# Collapse of the standard ferromagnetic domain structure in hybrid Co/Molecule bilayers

Mattia Benini (1), Andrei Shumilin (2), Rajib Kumar Rakshit (1), Antarjami Sahoo (1), Anita Halder (3,4), Andrea Droghetti (3), Francesco Cugini (5), Massimo Solzi (5), Diego Bisero (6), Patrizio Graziosi (1), Alberto Riminucci (1), Ilaria Bergenti (1), Manju Singh (1), Luca Gnoli (1), Samuele Sanna (7), Tomaz Mertelj (2), Viktor Kabanov (2), Stefano Sanvito (4), V. Alek Dediu (1)

(1) Istituto per lo Studio dei Materiali Nanostrutturati - CNR (ISMN-CNR), Via Piero Gobetti 101, Bologna 40129, Italy
(2) Department of Complex Matter, Jozef Stefan Institute, Jamova 39, 1000 Ljubljana, Slovenia
(3) School of Physics, AMBER and CRANN Institute, Trinity College, Dublin 2, Ireland
(4) Department of Physics, SRM University – AP, Amaravati 522 502, Andhra Pradesh, India
(5) Dept. Mathematical, Physical and Computer Sciences, University of Parma, Parco Area delle Scienze 7/A, 43124 Parma, Italy
(6) Department of Physics and Earth Science, University of Ferrara, Via Saragat 1, I-44122 Ferrara, Italy
(7) Department of Physics and Astronomy "A. Righi", University of Bologna, via Berti-Pichat 6/2, 40126 Bologna, Italy

**Abstract**

We show that, upon the chemisorption of organic molecules, Co thin films display a number of unique magnetic properties, including the giant magnetic hardening and the violation of the Rayleigh law in magnetization reversal. These novel properties originate from the modification of the surface magnetic anisotropy induced by the molecule/film interaction: the π-d molecule/metal hybridization modifies the orbital population of the associated cobalt atoms and induces an additional and randomly oriented local anisotropy. Strong effects arise when the induced surface anisotropy is correlated over scales of a few molecules, and particularly when the correlation length of the random anisotropy field is comparable to the characteristic exchange length. This leads to the collapse of the standard domain structure and to the emergency of a glassy-type ferromagnetic state, defined by blurred pseudo-domains intertwined by diffuse and irregular domain walls. The magnetization reversal in such state was predicted to include topological vortex-like structures, successfully measured in this research by magnetic-force microscopy. Our work shows how the strong electronic interaction of standard components, Co thin films and readily available molecules, can generate structures with remarkable new magnetic properties, and thus opens a new avenue for the design of tailored-on-demand magnetic composites.

**Introduction**

The microscopic understanding of the emergence of the macroscopic magnetic order is one of the earliest successes of quantum mechanics. In magnetism, Hund's coupling is responsible for the formation of the local magnetic moment of ions with partially filled shells, while the exchange interaction establishes how such moments align with respect to each other. The magnetic anisotropy then couples the lattice to the local moments, thus selecting their preferential direction of alignment. Remarkably, these three ingredients alone, local moment, exchange interaction and magnetic anisotropy, give rise to a multitude of magnetic states with sharply different static and dynamical properties. These go from conventional ferromagnets, to various types of antiferromagnets, to more complex magnetic textures such as vortexes and skyrmions[1]. Crucially, in conventional magnetism the exchange interaction and the magnetic anisotropy set the relevant length scales of a system and ultimately determine the dynamical response to an external perturbation.

We show both experimentally and theoretically that in hybrid Co/molecule bilayers, featuring well-known spinterface effects[2–7], the interplay between the magnetic exchange length and the correlation length of the random surface anisotropy term induced by metal-molecule orbital hybridization, leads to the formation of a glassy-type magnetic phase. This novel phase features properties radically different from those of the cobalt film alone, resulting in the colossal enhancement of the in-plane magnetic anisotropy, accompanied by the violation of the Rayleigh law at low magnetic fields and by the collapse of the standard domain structure. The magnetization in this state is characterised by blurred pseudo-domains intertwined by diffuse and irregular domain walls along with the perpendicular to the magnetization plane vortex-like topological defects.

The modification of the magnetic anisotropy MA of 3d metallic ferromagnetic films by the chemisorption of molecules on their surfaces has been recently reported in a number of communications[8–11]. Nevertheless, the theoretical modelling of such systems, was almost entirely based on Density Functional Theory (DFT) analysis of single molecule effects, giving a correct qualitative justification of the observed variations of the magnetic anisotropy[8,12,13] but failing to account for the numerical experimental values. Moreover, none of the theoretical reports has predicted any deviation from standard magnetic configurations, being limited to the framework of purely incremental effects.

In this work we also start from accurate DFT calculations on molecular scale and reveal an additional in-plane surface anisotropy term, where the variety in the different adsorption geometries effectively produces a random anisotropy field at the same length-scale. Next, we go beyond the single molecule approach and show for the first time that long-range correlations are crucial for the establishment of the extraordinary magnetic properties revealed in our experiments. This new finding opens conceptually new routes for the fabrication of novel materials with on-demand magnetic parameters.

Below we present a thorough experimental investigation of a variety of Co/Molecule thin film systems proceeding through temperature-dependent bulk magnetometry, hysteresis minor loops studies and magnetic force microscopy. Further our experimental findings are rationalised by a micromagnetic model, and by density-functional-theory (DFT) calculations, which provide the *ab initio* foundation of the model. A detailed discussion and comparison with the state-of-the-art knowledge concludes the paper.

**Experiment: Colossal enhancement of the coercive fields**

We investigate the modifications induced to the in-plane magnetic properties of Co thin films ($d$=3, 5, and 7 nm) interfaced with two different molecular species, namely fullerene, $C_{60}$, and Tris(8-hydroxyquinolinato)Gallium, $Gaq_3$. The Co/molecule samples are compared to two reference systems, consisting of aluminium-protected Co/Al and oxidized by air exposure $Co/CoO_x$. While the first represents one of the best ways to investigate ex-situ bare-like cobalt films[14], the naturally oxidized Co samples allow us to compare the features of the Co/molecule composites with those of a fully inorganic interface. In what follows we focus on 5 nm thick Co films, fully representative of all the claims advanced in this paper. Indeed, the behaviour of 3 and 5 nm samples is very similar with respect to the accuracy of our experiments, while the results on 7 nm samples were already partly reported[15].

We start from the most common characterization of a ferromagnetic material, namely the hysteresis loops, which are here measured by MOKE from room temperature (RT) down to 80 K. While the detailed temperature dependence will be discussed later, **Fig. 1** shows the results collected on the reference systems and on the two molecular-based cases at the intermediate and fully representative temperature of 150 K. The selected hysteresis loops of **Fig. 1a** clearly demonstrate the magnetic hardening induced by the molecular adsorption. The coercive fields, $H_C$, of Co/C$_{60}$ and Co/Gaq$_3$ are, respectively, 7-fold and 30-fold enhanced with respect to those of both the Al-capped and the CoO$_x$-interfaced Co thin films. The coercive-field enhancement produced in Co/C$_{60}$ is in qualitative and quantitative agreement with previously published data[9,11,16], indicating a strong hardening of the thin film. In contrast, the hybridization of the Co surface with Gaq$_3$ drives the hardening well beyond expectation, leading to a colossal enhancement of $H_C$. This persists even at room temperature, where the enhancement remains at about 100%. The coercive fields measured for 7 nm thick Co/molecule systems confirm the same trend but is characterized by a less pronounced hardening: at 80 K from 2- to 3-fold for Co/C$_{60}$ and Co/Gaq$_3$ respectively[15]. Noteworhily, while the hysteresis loop for Co/Gaq$_3$ in **Fig.1a** shows some imperfections with respect to regular hysteresis shape, these deviations are due to MOKE-induced minor arifacts, which do not replicate in SQUID measurements (see below) and even in most MOKE loops detected on similar samples (see more details in the SI). This allows us to exclude any significant presence of a second (or more) magnetic phase, which could have introduced errors in comparing experimental data with theory.

The hysteresis loops presented in **Fig. 1b** are those calculated on the basis of the micromagnetic model that will be introduced later on. For the moment, note that the model accounts for both the colossal $H_C$ enhancement and the significant broadening of the magnetic reversal transition along the loops, two features that are clearly detectable experimentally for Co/C$_{60}$ and further magnified in Co/Gaq$_3$.

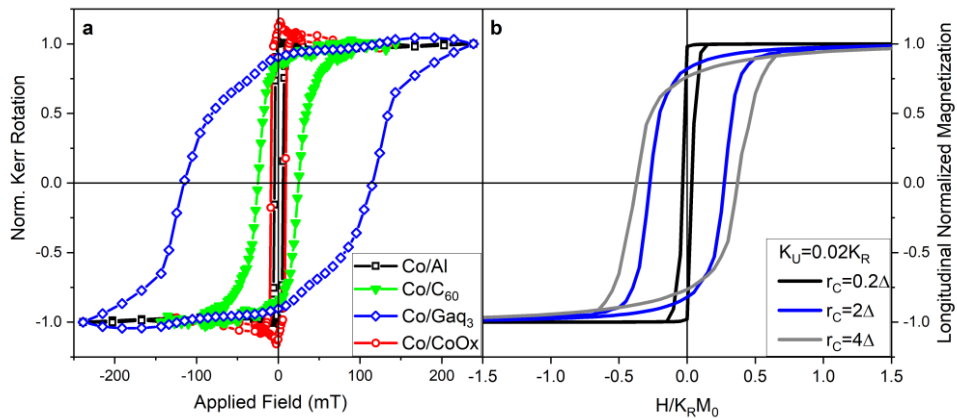

**Fig.1 a,** *Experimental hysteresis curves for Co/C$_{60}$, Co/Gaq$_3$ and reference Co/Al, Co/CoOx systems measured at 150 K, showing the broadening of the loops for Co interfaced with molecular species.* **b,** *Theoretical hysteresis curves obtained with the correlated random anisotropy model for $K_R=50K_U$ and different correlation radius values.*

### Experiment: Minor loops behaviour

Deeper insights into the observed magnetic hardening can be obtained from the hysteresis minor loops[17], as measured with MOKE (surface sensitive) and SQUID (bulk measurement) magnetometry. We start from a fully demagnetized state and proceed by subsequently increasing the field intervals of the magnetic loops, with the measurements taken at the representative temperature of 150 K for all the systems reported in **Fig. 1a**. While both techniques return very similar trends, SQUID measurements present more regular loops and allow us to measure the absolute values of the magnetic moment. As an illustrative example, in **Fig. 2a** we show the SQUID raw data for the minor loops of Co/C$_{60}$ (with subtracted background). All the other MOKE and SQUID data can be found in the

Supplementary Information (SI). The inset in **Fig. 2a** depicts a sketch of a typical minor loop and the parameters used for our quantitative analysis, namely the minor-loop coercive field, $H_C^*$, remanence $M_R^*$ and the maximum applied field $H_{MAX}$.

In **Fig. 2b** we report the relation between $H_C^*$ and $H_{MAX}$ for the reference and the molecule-interfaced cobalt layers, with data normalized to the average values of $H_C$ of the fully saturated loops (see SI for the non-normalized data). While the reference Co/Al samples are reproducibly characterized by a bisector-like growth at low fields, adsorbing molecules on Co leads to a radically different behaviour. This is characterised by a non-linear and very gradual increase of $H_C^*$ towards the fully saturated loop value at $H_{MAX} \gg H_C$. The deviation between the reference and molecule-based samples is further emphasised in the inset of **Fig. 2b**, where a distinct transition from a linear (reference samples) to a power-law behaviour (Co/molecule) indicates a drastic modification of the magnetization reversal mechanism at low magnetic fields.

We further analyse the minor loops data with the theoretical model of Takahashi et al.[18]. This is based on the well-known Rayleigh relation[19–21] describing the magnetization reversal via domain wall (DW) motion,

$$M = \chi H + \eta H^2, \qquad (1)$$

where $\chi$ is the initial susceptibility and $\eta$ is the Rayleigh constant associated to the Barkhausen jumps. Starting from this equation, the model defines four scaling power laws with associated critical exponents, relating various quantities of the minor loops. Among those one can derive a relation between the coercive fields and the remanent magnetizations,

$$\frac{H_C^*}{H_C} = \left(\frac{M_R^*}{M_R}\right)^n, \qquad (2)$$

where $n = 0.45$ is a universal exponent for all magnetic systems for which Eq. (1) is fulfilled. Such relation is typically satisfied by various ferromagnetic samples[22–24].

A log-log plot of the quantities entering Eq. (2) is provided in **Fig. 2c** and **Fig. 2d** for the reference and the Co/molecule samples, respectively. Note that data for both Co/Al and Co/CoO$_x$ are well compatible with the expected $n=0.45$ exponent. In particular, the behaviour of the oxidized sample is instructive, as it shows that the enhancement of the coercive field (in this case by surface oxidation) is not necessarily accompanied by a modification of Eq. (2) or more generally by any deviation from the Rayleigh law.

In contrast, when molecules are adsorbed on Co a significant deviation from the relation in Eq. (2) and hence from the Rayleigh law is observed. In fact, in **Fig. 2d** one can clearly see that the data for both molecular samples lie below the $n=0.45$ line, and even certainly deviate from a power law. While deviations from the Rayleigh-derived laws can be observed at relatively high fields, where the Rayleigh law does not properly account for the magnetisation dynamics[25], the Co/molecule systems are in disagreement with the expected trend for the whole field interval. This means that the magnetic behaviour of the molecule-interfaced cobalt thin films is not well described by the conventional magnetization reversal mechanisms of ferromagnets. While the observed magnetic hardening can in principle be naively explained by a mere rescaling of the anisotropy constants, the deviation from Rayleigh law indicates the establishment of a new magnetic state displaying an anomalous magnetization reversal mechanism. We will now theoretically analyse these new magnetic properties by first looking at how the molecules modify the surface anisotropy, and then by investigating how such modification affects the reversal mechanism.

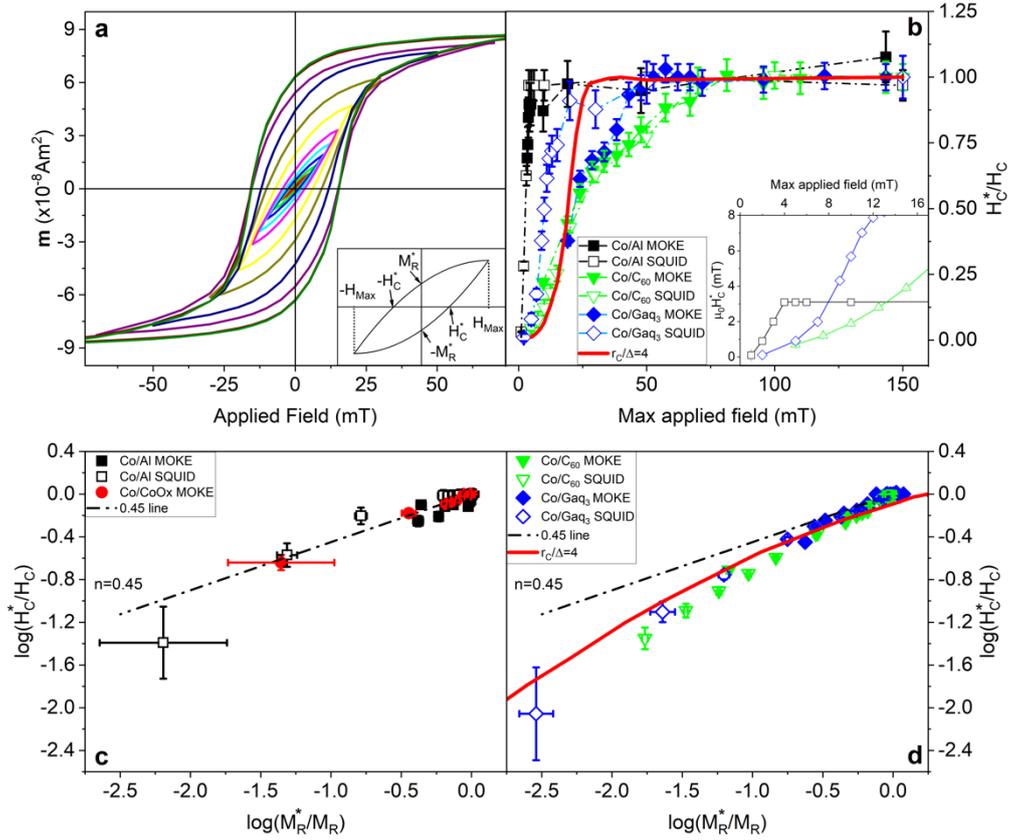

**Fig.2 a**, *example of minor loops obtained from SQUID magnetometry for a Co/C$_{60}$ system at T=150 K; inset shows a sketch of a minor loop along with the quantities of interest for the numerical fit.* **b**, *minor loops coercive fields normalized by the saturated value for the systems investigated; continuous red line is calculated within the correlated anisotropy model with r$_C$/Δ=4. Inset shows the low-field trend of H$_C$ for representative systems.* **c**, $\log(H_C^*/H_C)$ vs $\log(M_R^*/M_R)$ *trends for reference systems, well described by n=0.45.* **d**, *same plot for Co/Molecule systems showing a clear deviation from the standard behaviour, while the correlated anisotropy model with r$_C$/Δ=4 provides a reasonably good fitting (red line).*

### Theoretical investigation: first-principles calculations

We start our theoretical analysis by performing first-principles DFT calculations for a supercell containing a four-layer Co slab with either a C$_{60}$ or a Gaq$_3$ molecule adsorbed on the top surface (see **Fig. 3a,b).** We explicitly focus on the in-plane magneto-crystalline anisotropy (MCA) and, by definition, our frame of reference has the *z*-axis perpendicular to the surface, which lays in the *xy* plane. We consider FCC rather than HCP Co based on the recent tunnel-electron-microscopy images of our samples[26] and for simplicity we take the [001] surface. The Gaq$_3$ molecule is adsorbed with two O and one N atoms pointing toward the Co surface (**Fig. 3b**). After a geometry optimization, the angle in between the two quinoline ligands closer to the surface opens up until these ligands lay flat down on the surface, while the third ligand remains almost perpendicular to it[27,28]. For C$_{60}$ we consider two different adsorption geometries, labelled as 1 and 2, which differ by approximately a 45$^o$ molecular rotation (see **Fig. 3a**).

After the interface geometries are relaxed (see SI for more details), we estimate the MCA by rotating the Co magnetization in-plane and compute the system's energy $E(\theta)$ as a function of the magnetization angle, $\theta$, with respect to the *x*-axis. The direction of the minimum (maximum) energy corresponds to the slabs' easy (hard) axis, specified by the angle $\theta_{EASY}$ ($\theta_{HARD}$). The MCA energy is then defined as $E_{MCA} = E(\theta_{EASY}) - E(\theta_{HARD})$ and, by convention, the zero is set at the $\theta = 0°$ energy.

Note that the effect of the molecules on the MCA is purely electronic and it is induced by the hybridization of the molecular orbitals with the $d$ shell of the surface cobalt atoms, strongly modifying the orbital populations in the latter. All the magnetic effects are calculated on the Co slab and can be decomposed in atomic contributions, highlighting the surface character[29] (see SI).

**Fig. 3c** shows the effects that the molecules' adsorption has on the MCA. When compared to the clean Co slab, which is almost magnetically isotropic ($E_{MCA} \approx 0.1$ meV), a marked uniaxiality is induced by the presence of the molecules, resulting in an extraordinary enhancement of the surface MCA energy, up to 0.7 meV for $C_{60}$ and 1.5 meV for $Gaq_3$. Moreover, not only does the anisotropy increase, but the orientation of the easy axis also becomes strongly dependent on the adsorption geometry of the molecule. In the case of Co/$C_{60}$, $\theta_{EASY}$ switches from 120° to 40° when the molecule is changed from configuration 1 to 2 (see **Fig. 3c**), while for Co/$Gaq_3$, this modification is further enhanced, since the ligands are strongly chemisorbed on the surface in a very asymmetric fashion, and the molecule assumes a rod-like conformation. These magnetic effects are found to be rather local, meaning that they stem from the surface atoms directly bonded to the molecule, whereas the surface regions and the Co atoms, which are not covered by the molecule, remain almost unaffected.

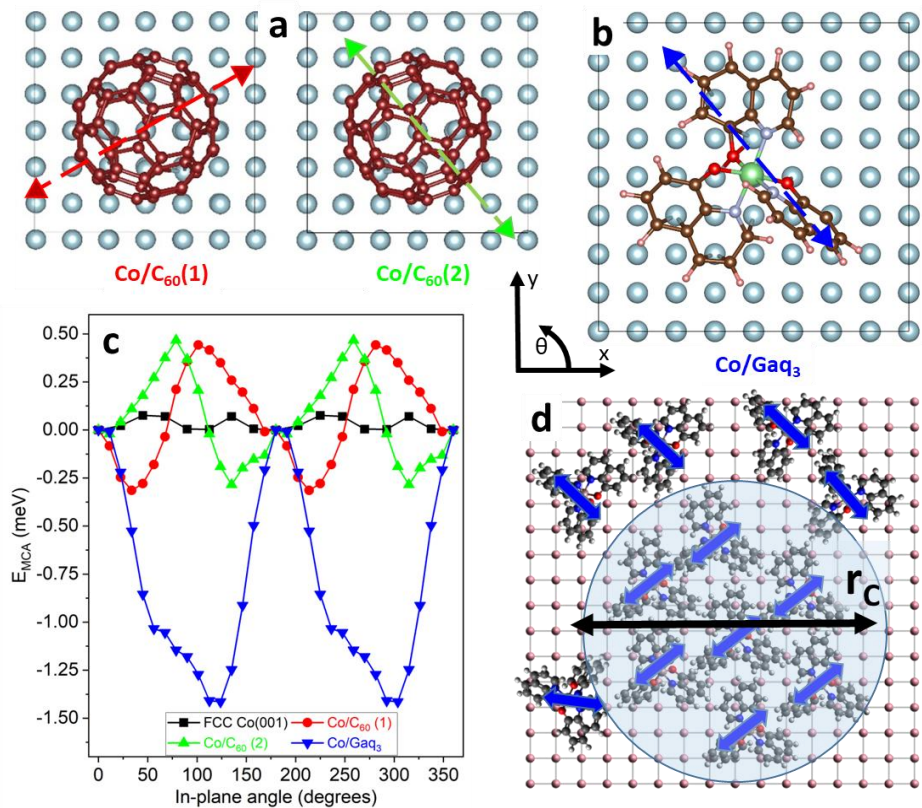

**Fig. 3** *Results of the DFT calculations. **a,** Top view of the two Co/$C_{60}$ adsorption geometries. **b,** Top view of the Co/$Gaq_3$ slab, alongside with the frame of reference highlighting the easy axis obtained from the DFT calculation. The Co atoms are in grey. The atoms of the molecules are respectively represented by brown (C), red (O), cyan (N), pink (H), and green (Ga) spheres* **c,** $E_{MCA}$ *as a function of the in-plane magnetization angle $\theta$ for two Co/$C_{60}$ and one Co/$Gaq_3$ slabs. The results are shifted so that $E(\theta=0°)=0$ meV. **d,** a schematic representation of the correlation radius $r_C$. The arrows on the different figures indicate the local anisotropy axes.*

In summary, the DFT calculations explicitly show that the surface-molecule hybridization redefines the MCA. In particular, it introduces strong symmetry modifications by creating a local, at the molecular scale, in-plane uniaxiality. The MCA orientation then depends on the chemisorption geometry. Since multiple molecular conformations are possible on the surface, the net result of the depositing molecules is that of creating a random anisotropy field with a typical lengthscale comparable to the size of the molecules themselves.

**Theoretical investigation: macroscopic model**

In order to describe the magnetization reversal dynamics we now move to micromagnetic model calculations. The described above local anisotropy encourages the application of the well-known empirical correlated random anisotropy model (RAM)[30,31]. The general concept of correlations in RAM models was used in a series of interesting works[32–34], but this approach was never employed for the description of such particular systems, consisting of fairly homogeneous magnetic thin films decorated by 2D randomly oriented islands of modified surface atoms. This expands the (calculated above) effect of the single molecule and distributes the local anisotropy over an area characterized by a certain correlation radius, $r_C$ (see **Fig. 3d**). The correlation radius $r_C$ is generated by molecular clusters with similar adsorption geometry and will be discussed in detail below.

Below we set the basis for the 2D Correlated Random Anisotropy Model (2D-CRAM), where the 2D limit indicates none or negligible variations of the magnetization along the axis perpendicular to the surface. This makes the model mostly adapted for the description of thin and ultra-thin magnetic films with random anisotropy.

We start from introducing the correlation term in the expression for the free energy:

$$\mathcal{F} = \frac{1}{2}\xi^2 \sum_\alpha (\nabla M_\alpha)^2 + K_Z(\boldsymbol{M} \cdot \boldsymbol{e}_z)^2 - K_R(\boldsymbol{M} \cdot \boldsymbol{a})^2 - K_U(\boldsymbol{M} \cdot \boldsymbol{e}_x)^2 - \boldsymbol{H} \cdot \boldsymbol{M}. \quad (3)$$

Here $\boldsymbol{H}$ is the external magnetic field applied in the *xy*-plane, $\boldsymbol{M}$ represents the magnetization density and $M_\alpha$ is its projection to the cartesian axis $\alpha$. The parameter $K_Z$ encapsulates a simplified description of the MCA and acts to keep the magnetization in-plane. Typically, this is stronger than other anisotropy terms. In the simplest approximation $K_Z \to \infty$, hence $\boldsymbol{M}$ is allowed to rotate only in plane, keeping constant the absolute magnetization value $M_0$ and explicitly shaping our model for the 2D limit. The first term on the right-hand side of Eq. (3) accounts for the exchange interaction. It includes the magnetic exchange length, $\xi$, that is usually of the order of a few nanometers[35], reflecting the energy enhancement in non-uniform magnetization scenarios. The $K_R(\boldsymbol{Ma})^2$ term is central to our model, since it represents the random anisotropy averaged over the film thickness. While the absolute value of $K_R$ remains constant, the coordinate-dependent easy axis $\boldsymbol{a(r)}$ is random. The distribution of $\boldsymbol{a(r)}$ is dependent on the correlation length $r_C$, and its interplay with the re-normalized magnetic length $\Delta = \xi/\sqrt{K_R}$ embodies the very core of the present model. The details on the correlation properties of $\boldsymbol{a(r)}$ with arbitrary $r_C$ are given in the SI. We also keep a small uniform anisotropy term $K_U$ defining the properties of the bare Co layer. This allows us to model Co samples without molecular overlayers by simply setting $K_R=0$. We do not explicitly consider dipole-dipole interaction since it is suppressed in thin films with in-plane magnetization and we assume it to be irrelevant compared to the random anisotropy.

Considering the possibility to realize the long-scale correlation, we note that the growth of molecular layers on metallic or oxide surfaces typically involves tens of nm large nucleation islands[36,37]. The growth process is characterized by a finite diffusion time of the molecules on the surface, which allows them to bind in preferential, lowest energy configurations.

To find the magnetic properties of the 2D-CRAM, we perform numerical simulations based on the minimization of $\mathcal{F}$ and the Landau-Lifshitz-Gilbert equations. The analysis is performed considering only two key parameters: the dimensionless correlation length $r_C/\Delta$ and the ratio $K_U/K_R$, which describes the relation between uniform and random anisotropy. For details see the SI.

The first significant result is reported in **Fig. 1b**, which shows the dependency of the calculated hysteresis loops on $r_C$ at a fixed $K_U/K_R$ ratio (see SI for different $K_U/K_R$). For low correlation lengths the film properties are controlled by $K_U$, even when the random anisotropy is 50 times larger than the uniform one (see $K_u=0.02K_R$, $r_C=0.2\Delta$ curve in **Fig. 1b**). In this case the coercive field is low and the shape of hysteresis loop is almost square, similarly to what is observed experimentally for the reference Co/Al film in **Fig. 1a**. When $r_C$ is increased to values comparable to or larger than $\Delta$ the effects of the random anisotropy start to prevail: the hysteresis loops broaden, resulting in a strongly increased coercive field. Moreover, the shape also changes, displaying a broader, more gradual, magnetization reversal, similar to that of Co films interfaced with molecules in **Fig. 1a**. In the SI we show that, if the hysteresis loops for Co/C$_{60}$ and Co/Al samples are rescaled by setting the applied field in units of the closure field of Co/C$_{60}$, the data coincide with the model loops obtained for $K_U=0.02K_R$ with $r_C/\Delta=4$ and $r_C/\Delta=0.2$, respectively.

In addition to the hysteresis loops, our model fairly accounts for the modification of the minor loops of the Co/molecule systems. It can be seen in **Fig. 2** that the model adequately describes the minor loops evolution with field, capturing the non-linearity and the steady enhancement of the coercive field $H_C^*$ (**Fig. 2b**). Moreover, the application of the model describes very well the deviation from the power law in **Fig. 2d**.

Remarkably, our detailed micromagnetic simulations reveal an unusual glassy-type magnetic configuration for high correlation lengths. Even though in agreement with previous theoretical predictions[34,38] experimental evidences for such magnetic phase have never been reported before. **Fig. 4** shows the maps of the normalized magnetization components $m_\alpha = M_\alpha/M_0$ with and without the random anisotropy term. The ferromagnetic film without a molecular overlayer is described by $K_R=0$ and $K_U=1$ (**Fig. 4a,b** - in this case $\Delta$ is defined as $\xi/\sqrt{K_U}$), and a standard magnetic configuration is clearly observed in its pseudo ground state, namely in the fully demagnetized state. For the magnetic field aligned along the $x$ axis the division in domains oriented along the field (yellow) and opposite to it (blue) is well defined and the two are separated by narrow domain walls, with the presence of Bloch points. This magnetic configuration is standard for thin ferromagnetic films and the calculated domain wall width of 10-20 nm is in good agreement with the experimental data[35].

This domain configuration clearly collapses for finite $K_R$ and $r_c > \Delta$. For example, for $K_U = 0.02K_R$ and $r_c = 4\Delta$ a totally different magnetic configuration develops (**Fig. 4c,d**). It is characterized by a gradual transition between irregular islands of opposite magnetic orientations, separated by broad and diffuse transition regions (pseudo domain walls). The structures in Fig.4c-e strongly depend on the cooling procedure, described in detail in the SI part.

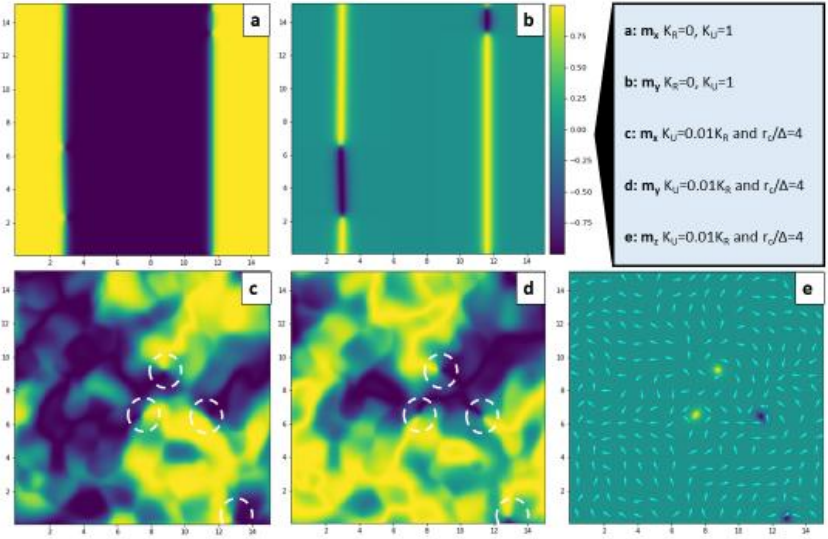

*Fig. 4* Configuration of the normalized magnetization vector *m* components in the system's fully demagnetized state. Yellow (blue) indicates $m_\alpha = \pm 1$. **a**, $m_x$ and **b**, $m_y$ for the case $K_R=0$ and $K_Z=2K_U$ (i.e. no random anisotropy term): uniform anisotropy term dominates, and a standard domain configuration is achieved as expected. When the film properties are controlled by random anisotropy (**c,d**) the domain structure collapses and *m* continuously varies in random directions (depicted with white arrows). **e**, For $K_R$ comparable to the shape anisotropy (in this case, $K_Z=2K_R$) localized vortexes with finite z-component appear, their position being highlighted in **c,d** by white dashed circles.

Noteworthily, we confirm that the glassy state posses also an additional and important characteristic fingerprint, i.e. the presence of topological defects[34,38], represented in our systems by magnetic vortexes with an out-of-plane orientation of the magnetization. This is found to happen at the singularity-like points in the *x-y* plane, where pseudo-domains with opposite orientations come in direct contact with each other (circles in **Fig. 4d**). The magnetization has a $2\pi$ rotation in the loop around this point, as shown in **Fig. 4c-e**. In addition, the predicted vortexes are evidenced by the *z*-component of the magnetization in **Fig. 4e** and in **Fig. 5 a-c**. Along a hysteresis loop, the orientation of the vortexes magnetization depends on the magnetization history. For the virgin demagnetized state and during the initial magnetization both vortexes and anti-vortexes are present (**Fig. 5a,d**). However, considering that an inevitable small angle tilting of the magnetic field with respect to the film plane is present (**Fig. 5d**), during the field reversal for the positive branch of the loop all vortexes are oriented along the indicated perpendicular-to-plane component (**Fig. 5b**) and switch to the opposite direction at the reversed field (**Fig. 5c**).

Remarkably, high-resolution magnetic-force microscopy (MFM) images of the prototypical Co(5nm)/$C_{60}$(25nm) samples reveal a well-established system of vortexes at room temperature. While the detection of vortexes on the virgin magnetization curve has proven extremely challenging, we focused on the left and right hysteresis branches and performed the magnetic imaging at a small field of 2 mT applied after fully saturating the magnetisation in the opposite direction. One can clearly see in **Fig. 5e** a number of vortex-like black magnetic defects, located in morphologically smooth areas (see the details for all the observed vortexes and respective morphologies in SI). Zooming on a single vortex (**Fig. 5f**) reveals a typical size of 50-100 nm, an accuracy, which cannot be exceeded in MFM studies. All vortexes are oriented in one direction (black colour) in perfect agreement with magnetization history in **Fig. 5d**. The successful detection of vortexes by MFM imaging (**Fig. 5e,f**) represents a strong verification for the existence of this unusual magnetic phase and it will be discussed in more details below. Note that the in-plane magnetization part cannot be detected by MFM, and it is generally extremely challenging considering the nanometer characteristic scales.

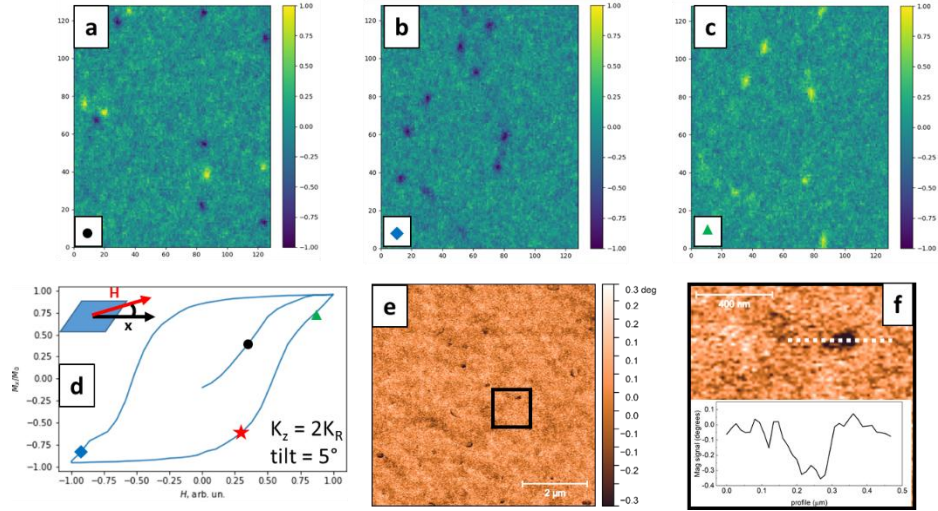

***Fig.5 a,b,c,*** *Micromagnetic simulation of the spatial distribution of $m_z$ calculated at different representative points of a simulated hysteresis loop with an applied field tilted 5° out of plane (magnetization indicated in **d** by the circle, rhomb and triangle). Different vortexes configurations are predicted depending on the magnetization history. The MFM imaging of vortexes and their spatial distribution in the Co(5nm)/C$_{60}$(25nm) sample at RT and under an applied field of 2 mT is shown in **e**, where several black spots with lateral dimension of about 50 nm. The black square is shown zoomed in **f** for a better visualization of a representative dark spot, along with its lateral profile. The images in **e** and **f** were taken at the magnetization indicated by red star in **d**.*

**Experiment: temperature dependence of the coercivity**

Finally, we discuss the temperature dependence of the hysteresis loops. **Fig. 6** shows the extraordinary enhancement of the coercive fields with reducing the temperature from RT down to 80 K. This trend is particularly clear when compared to both reference samples, and it is in reasonable agreement with the Jiles-Atherton model[39], in line with previously reported trends measured on Co/C$_{60}$[9]. The exponential dependence of the coercivity with temperature is clearly visible in **Fig. 6b** and indicates no conceptual modification of the temperature behaviour of coercivity with respect to the standard ferromagnetic state. Our data show no additional arguments either in favour or against the interesting hypothesis[16] of the significant modification of the metal-molecule hybridization at about 150K due to the freezing of the rotational/vibrational molecular degrees of freedom. Nevertheless, the detection of vortexes at room temperature allows us to claim that in Co/Molecule systems the Glassy magnetic phase is present both below and above the possible freezing temperature.

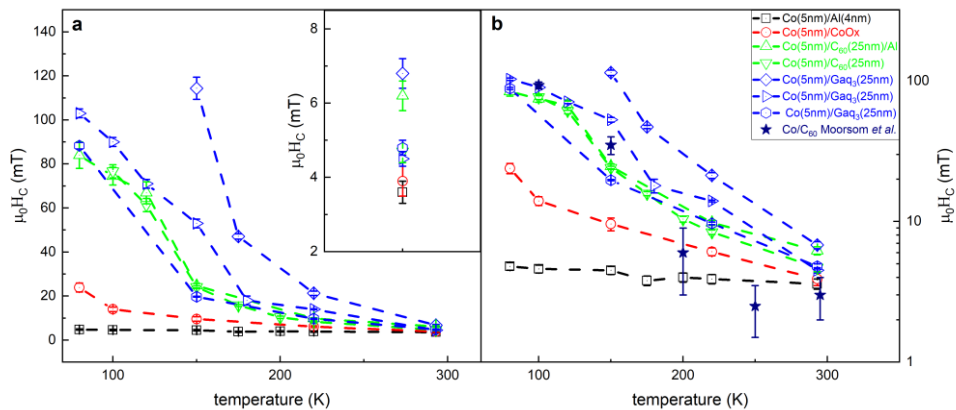

***Fig.6 a,*** *The coercive field as a function of temperature for Co/C$_{60}$ and Co/Gaq$_3$, compared with the reference Co/Al and Co/CoOx systems. Inset: enlargement of the values of $\mu_0 H_C$ at 293 K. **b** same dataset reported in log-lin form, along with Co/C$_{60}$ SQUID data presented in Moorsom et al., 2014[9], in reasonably good agreement with our experimental data. The authors thanks Dr. Timothy Moorsom (University of Leeds) for sharing the data.*

**Discussion**

Currently, magnetism and spintronics have been significantly energised by the rise of numerous types of novel, and sometimes exotic, materials, such as topological insulators, 2D magnets, altermagnets etc. Yet, artificial compounds, where the conventional properties of their components are combined to display an anomalous behaviour of the composite, represent a vast playground for magnetic materials with tailored-on-demand properties. We have here presented evidences that the hybridization of Co thin films with various molecules leads to the formation of a glassy magnetic phase, displaying a significantly enhanced magnetic anisotropy along with the violation of the Rayleigh relation at low fields, and other unusual properties.

Such a glassy state has been theoretically predicted time ago, but it has been never observed experimentally. For example, Chudnovsky and colleagues[32,38] anticipated a new magnetic phase, called Correlated Spin Glass (CSG) and characterized by multiple equilibria. They defined its order parameter and a number of uncommon properties, among which the lack of domain walls between Imry-Ma type domains[40], topological defects and other. In the previous paper[32] it is claimed that CGS has no magnetic remanence and no coercive field, while in the more recent paper it is shown that the magnetic order depends on the cooling-heating history and finite remanence is likely to appear in realistic experimental conditions. Indeed, the zero-remanence requires a specific cooling of a demagnetized system[32] and its realization is experimentally very challenging. A similar model was described by Dieny and colleagues[34] where the glassy shape of Imry-Ma domains is for the first time projected theoretically.

Our findings are in great agreement with these predictions and represent their first experimental realization. It is not surprising perhaps that this predicted and unobserved behaviour was found by us in systems radically different from those to which RAM was routinely applied (amorphous[25,30,41] or nanocrystalline FM[42], magnetic alloys with impurities[43] etc.). Moreover our work shows that this glassy phase does not represent a purely academic interest, its properties are greatly appealing for various magnetic applications. The hardening effect presented in this paper and even stronger hardening communicated recently for similar systems[44] represent a novel and strongly versatile route for designing and building RE-free hard magnetic materials with on-demand parameters.

Scientifically, our research highlights the crucial importance of the correlation effects and respective length scales for the comprehensive understanding of the modifications of the magnetic properties in ferromagnetic thin films hybridized with molecules. Previous explanations of the $H_C$ enhancement were merely based on DFT calculations performed at the atomic/molecular scale. While able to justify qualitatively the magneto-crystalline enhancement, this approach could neither fit the hysteresis loops nor predict the violation of the Rayleigh law or other unusual macroscopic features.

Indeed, it is the formation of the glassy state which gives rise to this plethora of novel properties, from the hardening to the unreported previously violation of the Rayleigh law, from the topological defects to the collapse of the domain structure etc. The previously proposed name of CSG in our opinion is partly misleading and was perhaps inspired by the first calculations showing zero-remanence. We believe that the correlated ferromagnetic glass (CFG) could be a more appropriate term to describe this phase (as also proposed in[45]), but the correct terminology can be defined only in further debates and discussions.

Note that the CSG or CFG and all the related effects emerge only for $r_C/\Delta > 1$. Considering that typical exchange length for thin cobalt films are in the range of 15-20 nm (our calculations and the FCC case in *Vaz et al.*[35]), and taking into account the definition $\Delta$ ($\Delta = \xi/\sqrt{K_R}$), for typical coercive field $H_C$ of 100 mT and dimensionless $K_R = 1.4$, the ratio $r_C/\Delta > 1$ is obtained for $r_C > 10\text{-}15$ nm or equivalently 10-15 molecules bound in a correlated mode. The glassy phase is realized in the regime of intermediate thicknesses (approximately 3-10 nm), that is above the well known in-plane to out-of-plane magnetization switching instability[10], and below thicknesses where the bulk properties strongly dominate the magnetization dynamics. Note, that in a previous paper we demonstrated that the interface-induced magnetic hardening propagates to length scales of several nanometers[15].

We assign the microscopic origin of the glassy phase to the interplay between the anisotropy correlation length and the exchange length, both competing to define the shortest scale for the magnetisation rotation and introducing an instability, which leads to the collapse of the standard domain structure. Similar interplay of competing lengths is fundamental in superconducting materials (competition between coherence length and penetration depth) or for metal-insulator phase transitions (competition between Bohr and screening radii).

In conclusion, our research demonstrates the emergence of the glassy ferromagnetic phase in a simple and easy to reproduce system. Its appearance is caused by a random but correlated anisotropy field established at the interface between metallic and molecular components for long range correlations. The Corelated Spin Glass (or Correlated Ferromagnetic Glass, as proposed in this paper) features rich and complex new physics, calling for thorough further research, but it also represents a promising and versatile technology for the creation of conceptually innovative high-anisotropy magnetic materials.

# METHODS

**Samples fabrication**

Co/Molecule and Co/Al bilayers were grown in UHV condition on $Al_2O_3$(0001) single-crystal substrates. The Co deposition was done by means of the electron beam evaporation (base pressure of $2.5 \times 10^{-9}$ mbar, rate 0.03 Å sec$^{-1}$) with the substrate kept at RT to promote the polycrystalline growth, confirmed in the previous publication[15]. The overlayers ($C_{60}$, $Gaq_3$, Al) were subsequently deposited, without breaking the vacuum, by the thermal evaporation from effusion cells (base pressure of $1 \times 10^{-8}$ mbar) and the substrate kept at RT. Deposition rate: 0.15 Å sec$^{-1}$ for $C_{60}$, 0.25 Å sec$^{-1}$ for $Gaq_3$ and 0.1 Å sec$^{-1}$ for Al.

**Magneto-optical characterization**

Hysteresis loops were recorded with an L-MOKE setup, the light wavelength $\lambda = 638.2$ nm (He-Ne laser). Samples were put in a cryostat and measured in a pressure of $10^{-6}$ mbar in vacuum condition. All temperature-dependent measurements were performed by cooling the systems in zero applied field. The hysteresis loops were subject to a symmetrisation procedure in order to remove quadratic Magneto-Optical signals. Hysteresis loops were taken with applied field varied in steps, with an effective field sweep rate of the order 1 mT/sec.

**SQUID characterization**

The magnetic moment of Co/$C_{60}$, Co/$Gaq_3$ and Co/Al bilayers was measured using the DC option of a Quantum Design superconducting quantum interference device (SQUID) magnetometer MPMS-XL5. Hysteresis loops were recorded at 150 K with the magnetic field applied parallel to the film surface. The samples were first cooled with a zero applied magnetic field, and then a series of minor loops were recorded increasing the maximum applied field.

**MFM characterization**

In-field MFM images were recorded using the phase detection mode, i.e., monitoring the cantilever's phase of oscillation while the magnetic tip was scanning the sample surface at a distance of 50 nm (lift mode). In order to exclude the influence of the tip on the magnetic state of the sample, we used different scanning directions and tip to sample distances, obtaining the same results with different operating conditions.

**DFT calculations**

The Co surfaces are modelled using periodically repeated slabs with a square surface supercell of (4× 4) periodicity and four layers thickness. The molecules are adsorbed on one of the slab's surfaces. The calculations are spin polarized. The Perdew-Burke-Ernzerhof (PBE) generalized gradient approximation (GGA)[46] The geometry relaxations are carried using the Fritz Haber Institute ab-initio molecular simulations (FHI-aims) all-electron code[47] is selected for the exchange-correlation functional. A standard numerical atom-centred orbitals basis sets "tier 2" and a 3x3x1 **k**-point mesh is employed. The atomic positions of two bottom layers of the slabs are maintained fixed, while all other atoms are relaxed until the ionic force are smaller than 0.01 eV/Å.

After relaxation, the magnetic properties are predicted using the projector augmented wave (PAW) method[48] as implemented in the Vienna Ab-initio Simulation Package (VASP)[49] with the PBE exchange correlation functional. The tetrahedron integration method with a kinetic-energy cutoff of 600 eV is employed. An energy convergence criterion equal to $10^{-7}$ eV for total energy calculations is adopted. The **k**-point sampling is performed using a MonkhorstPack (MP) grid with 12×12 **k**-point mesh in the two-dimensional Brillouin zone. The energies $E(\theta)$ are obtained by means of the magnetic force theorem[50,51] followings two steps. Firstly, a scalar relativistic collinear charge self-consistent calculation is carried out to obtain the charge density. Then, that charge density is used as input in noncollinear calculations performed non-self-consistently including SOC, where the magnetization vector is oriented along different directions, and $E(\theta)$ is approximated as the band energy. The tetrahedron integration method with a kinetic-energy cutoff of 600 eV is employed. An energy convergence criterion equal to $10^{-7}$ eV for total energy calculations is adopted. The atomic SOC energies are defined as $E_{SOC} \sim \langle \frac{1}{r}\frac{dV}{dr} \boldsymbol{L} \cdot \boldsymbol{S} \rangle$ where $V(r)$ is the spherical part of the effective potential within the PAW sphere, and $\boldsymbol{L}$ and $\boldsymbol{S}$ are orbital and spin operators, respectively. These expectation values are about twice the actual values of the total energy correction to the second order in SOC[52].

**Acknowledgements**

M. Benini, A. Shumilin, R. K. Rakshit, A. Halder, P. Graziosi, A. Riminucci, M. Singh, V. Kabanov, T. Mertelj, A. Drogetti, S. Sanvito and V. A. Dediu acknowledge the support of the EC project INTERFAST (H2020-FET-OPEN-965046). L. Gnoli and I. Bergenti acknowledge the support of the EC project SINFONIA (H2020-FET-OPEN-964396). A. Halder was supported by European Commission through the Marie Skłodowska-Curie individual fellowship VOLTEMAG-101065605. A. Sahoo acknowledges the International Centre for Theoretical Physics for supporting the ICTP-TRIL fellowship.

**Supplementary information for: Collapse of the standard ferromagnetic domain structure in hybrid Co/Molecule bilayers**

**S.I AMR DATA**

**Fig. S1** shows the AMR values measured at 80K for the Co(5nm)/Al(4nm) and for Co(5nm)/Gaq$_3$(25nm) in both transverse (**H** in plane and perpendicular to **I**) and longitudinal (**H** in plane and parallel to **I**), with an applied voltage of 100 mV. The coercivity is extracted from the data by locating the H value corresponding to the AMR peaks. The Co/Gaq$_3$ sample has a value of (130 ± 1) mT while a value of (5 ± 1) mT is found for the reference Co/Al, in agreement with MOKE measurements performed at the same temperature.

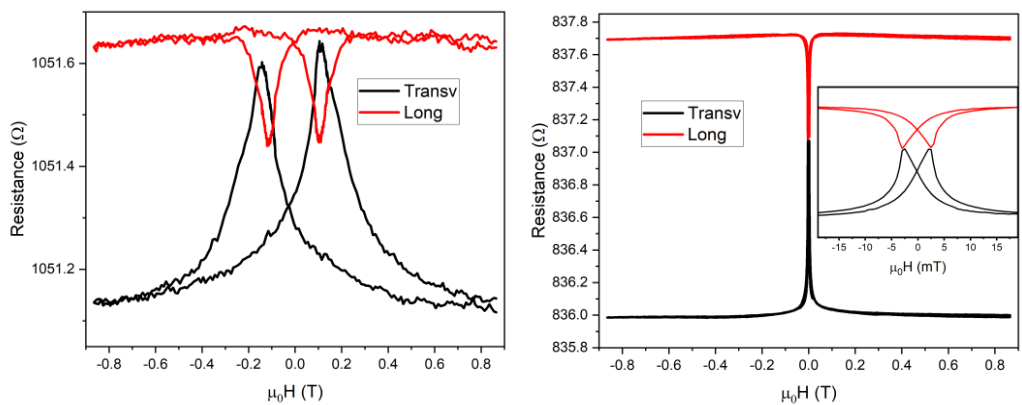

**Fig. S1** *AMR data for Co/Gaq$_3$ sample (left) and reference Co/Al (right) obtained at 80 K in both transverse (black lines) and longitudinal (red lines).*

**S.II MINOR LOOPS CHARACTERIZATION**

The minor loops obtained by MOKE magnetometry at 150K are reported in **Fig. S2**, while the ones obtained by SQUID magnetometry are reported in **Fig. S3**. Note that the loops clearly indicate the absence of any detectable second magnetic phase. The non-normalized plot of coercive field vs maximum applied field (data used for **Fig. 2b** in the man article) is displayed in **Fig. S4**, with added data for Co/CoOx for completeness. It is worth noting that the values extracted from Co/Gaq$_3$ and Co/C$_{60}$ by SQUID measurements are lower in absolute value with respect to the ones obtained by MOKE measurements. This mismatch is attributed to SQUID measurements not being done with the applied field along the samples easy axis, thus resulting in a lowering of the measured coercivity.

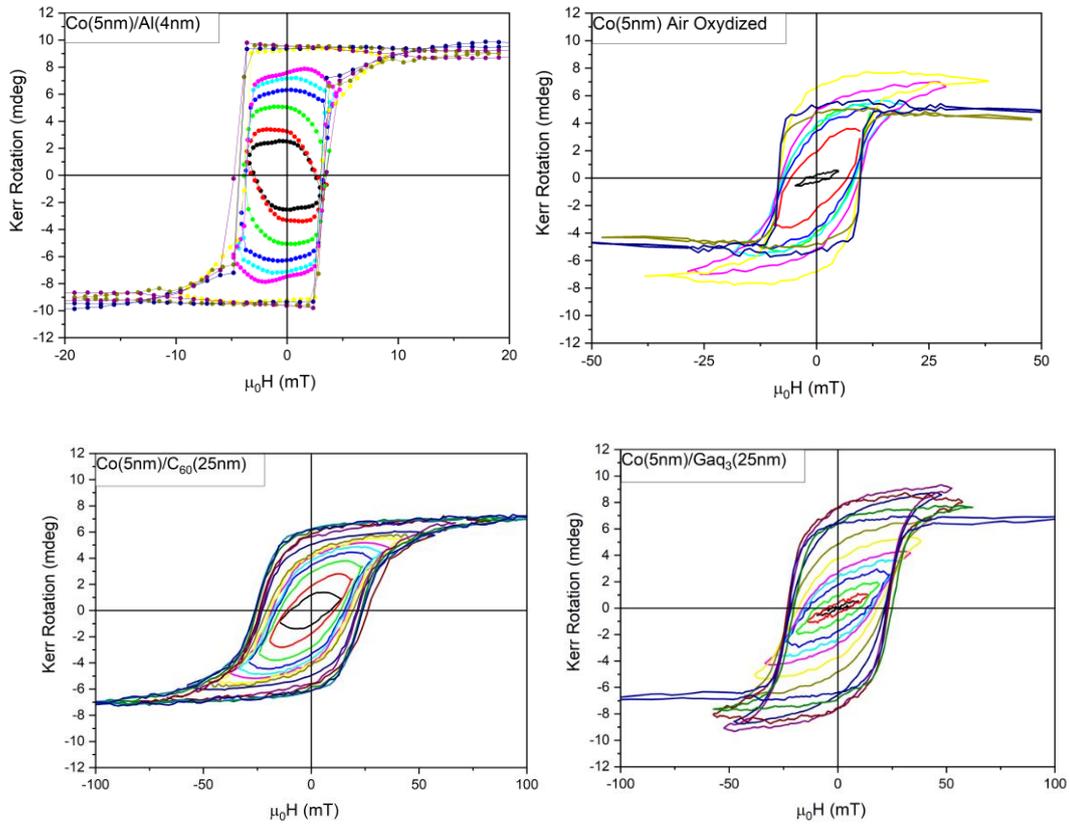

**Fig. S2** *Minor loops at 150K for reference Co/Al (top-left) Co/CoOx (top-right). Co/$C_{60}$ (bottom-left) and Co/Gaq$_3$ (bottom-right). The x-scale is varied for each sample for better evidence of the smaller loops.*

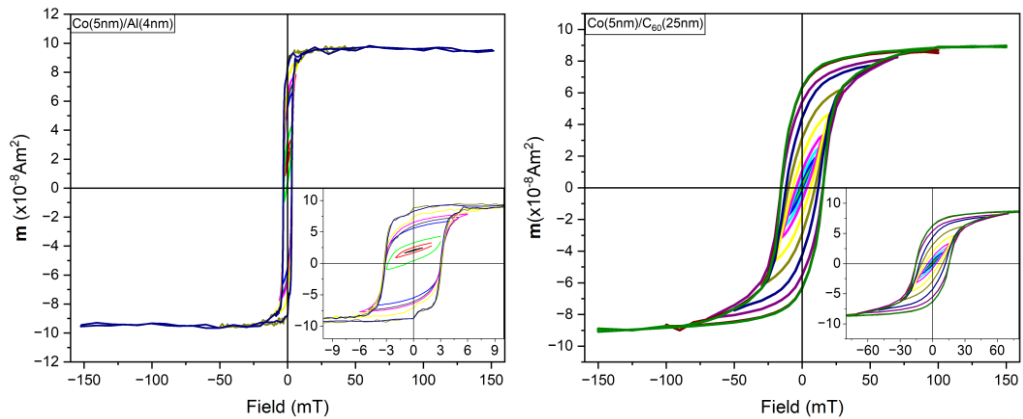

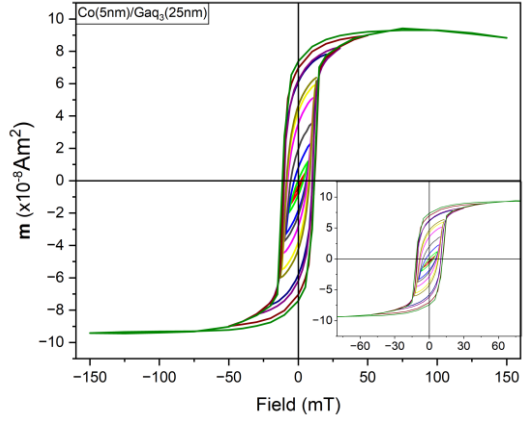

**Fig. S3** *Minor loops for the Co/C$_{60}$, Co/Gaq$_3$ and reference Co/Al obtained by SQUID magnetometry at T=150K. Insets show the zoomed-in central part of the graphs for better evidence of the smaller loops (quantities and units of x and y axis are the corresponding of the non-zoomed graph).*

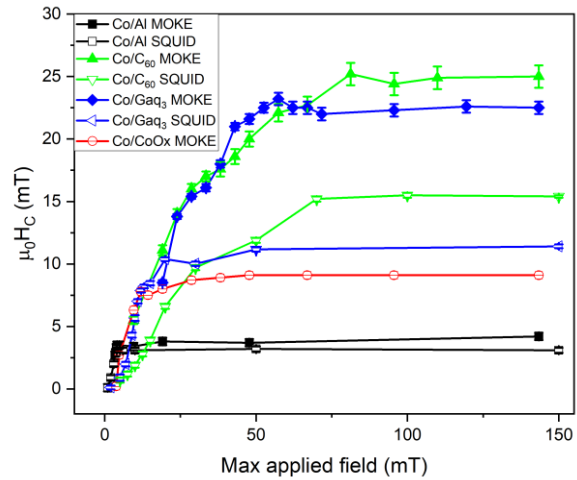

**Fig. S4** *Absolute values of the coercivity as a function of the minimum applied field in each loop.*

**S.III DFT CALCULATIONS, CALCULATED ATOM RESOLVED ANISOTROPY**

We demonstrate here that the enhancement of E$_{MCA}$ predicted by DFT in Co/Gaq$_3$ is an interface effect induced by the molecule. We resolve the contribution of each atom *i* to $E_{MCA}$ by computing the difference of the SOC energy per atom, $E_{MCA}(i) \propto \Delta E_{SOC}(i) = E_{SOC}(i, \theta_{EASY}) - E_{SOC}(i, \theta_{HARD})$. Negative (positive) values of $\Delta E_{SOC}(i)$ indicate contributions that increase (reduce) the slab's MCA. As shown in **Fig. S5**, $\Delta E_{SOC}(i)$ fluctuates. Nevertheless, it has the largest value for a few surface atoms, which are mostly those forming a covalent bond with the molecule. Hence, these atoms contribute the most to the MCA. This result can be highlighted by carrying out a partial summation $S(N') = \sum_{i=1}^{N'} \Delta E_{SOC}(i)$ with $N' \leq N$. $S(N')$ is rather small when considering only the atoms in the bottom layers of the slab, while it jumps of an order of magnitude when reaching interface layer.

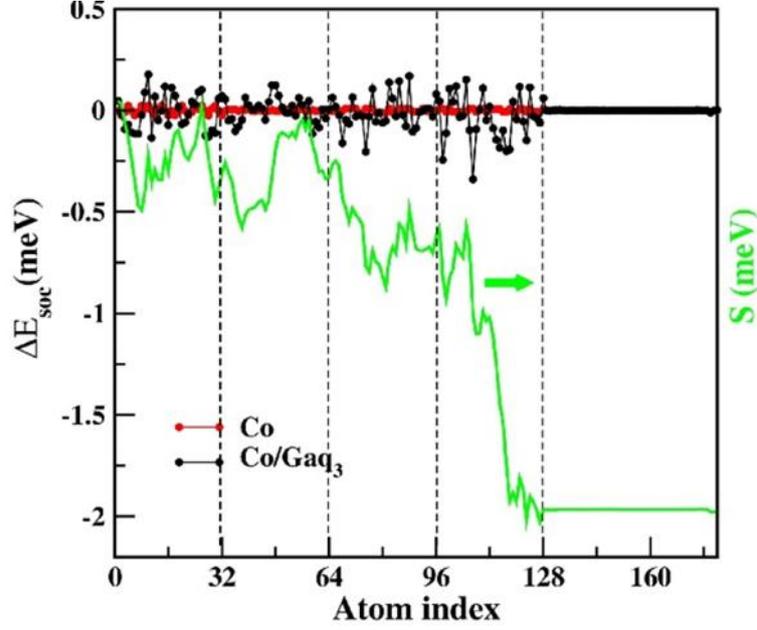

**Fig S.5** *SOC energy difference $\Delta E_{SOC}(i) = E_{SOC}(i, \theta_{EASY}) - E_{SOC}(i, \theta_{HARD})$ per atom i=1, ...,180. Red and Black points are respectively for the clean Co slab and the Co/Gaq$_3$ slab. Atoms labelled from 96 to 128 are at top surface layer, where the molecule is adsorbed, whereas atoms from 0 to 32 are at the bottom surface. Atoms labelled from 128 to 180 are the atoms of the molecule. The green curve is the cumulative sum $S(N') = \sum_{i=1}^{N'} \Delta E_{SOC}(i)$ used to highlight that the largest contribution to the MCA comes from the interface atoms.*

**S.IV MACROSCOPIC-SCALE CALCULATIONS**

To study the model described by Eq. (1) in the main text, we introduce the discrete lattice that can be considered as its mapping to the "classical spin glass" model[1,2]. It allows us to use a variety of conventional numerical methods for its study. However, the mapping implies that the anisotropies on neighbour sites are strongly correlated. The step of the lattice should be small compared to $\Delta$, while the size of the whole lattice should be large compared to the features of magnetic state. It limits the values of $r_c$ that are possible to simulate with our methods. All the results present in the article correspond to 128x128 lattice with size L=60Δ. They are checked to be robust against modification of lattice size and step.

The calculations start with the generation of random but correlated anisotropy field that was achieved with NAG library. First, we generate two random fields $a_x$ and $a_y$ with the Gaussian variogram using the NAG routine g05zrf with correlation length $r_c$. The distribution of the random angle f was calculated as:

$$\phi = \arctan\left(\frac{a_y}{a_x}\right). \quad (S1)$$

As a result, the random field f is correlated with the radius $r_c$. A typical distribution of the anisotropy angles is presented in **Fig. S6 (b)**. The correlation function of the anisotropy angles is plotted in **Fig. S6(a)**. It is clear that the spatial distribution of the anisotropy angles is not Gaussian but the correlation function is decreasing on the scale of $r_c$.

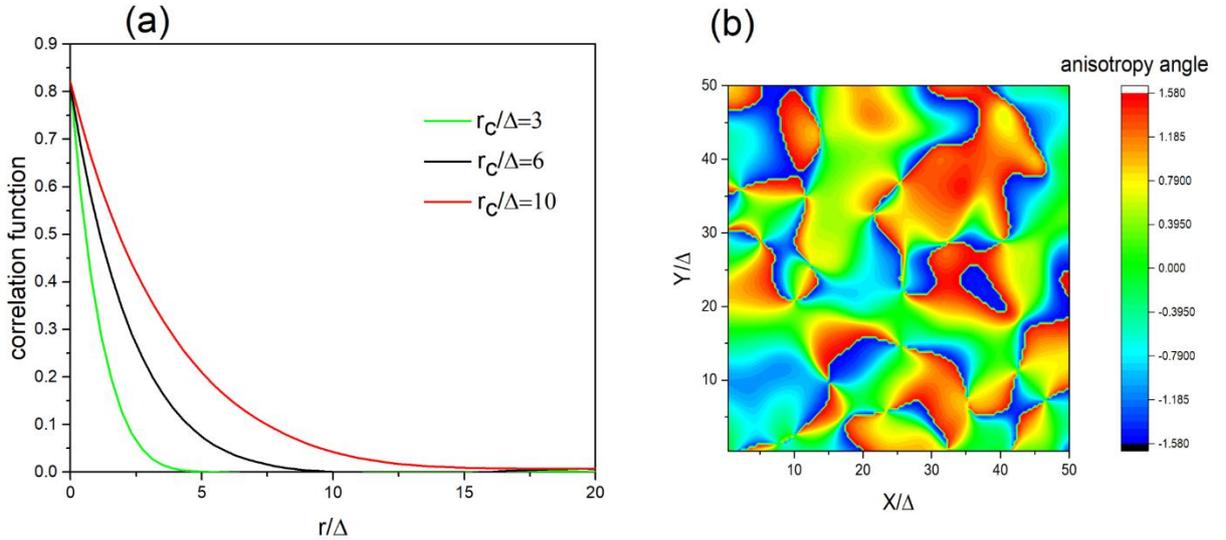

**Fig. S6 (a)** *Dependence of the angle correlation function on distance.* **(b)** *Distribution of the anisotropy field over the sample for $r_c/\Delta = 4$.*

The magnetic states are found by the minimization of the free energy integrated over the sample surface.

$$\int \mathcal{F} d^3 r \to min \quad (S2)$$

that is done numerically with the conjugate gradient method.

The ferromagnetic material described with Eq. (1) in the main text has many local minima of free energy at small magnetic fields. Therefore, the local minimum found with the numeric procedure depends on the initial state of the system. When the external field $\boldsymbol{H}$ is changed adiabatically by the small value ($\boldsymbol{H} \rightarrow \boldsymbol{H} + \delta \boldsymbol{H}$) the minimum for previous field $\boldsymbol{H}$ can be used as an initial state to the modified field $\boldsymbol{H} + \delta \boldsymbol{H}$. For the calculation of the full hysteresis loops the procedure starts from a high magnetic field $H \gtrsim K_R M_0$ and the uniform saturated magnetic state, $\boldsymbol{M} = M_0 \boldsymbol{e}_x$.

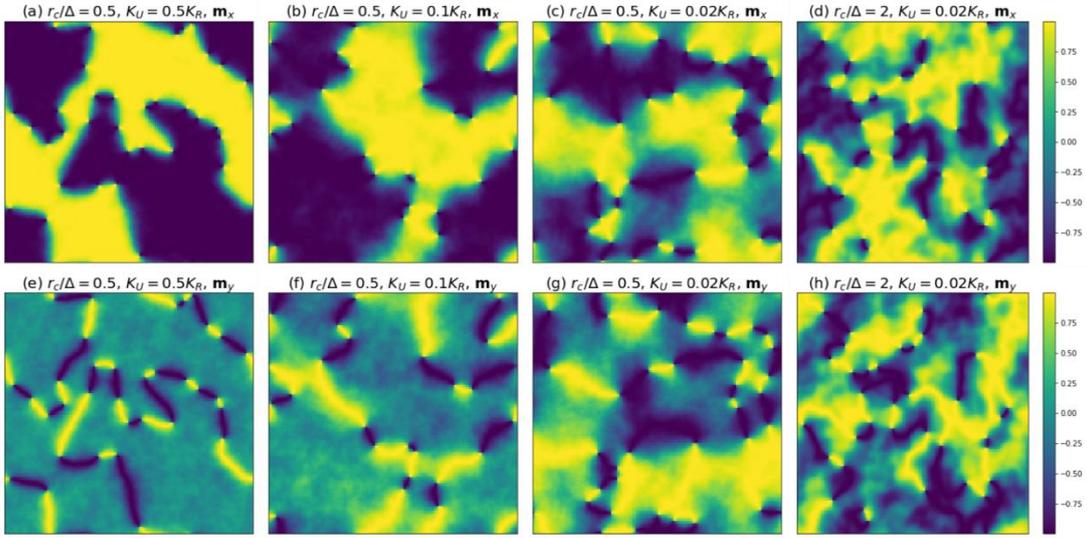

**Fig. S7** *Demagnetized states for different parameters* $r_c/\Delta$ *and* $K_U/K_R$.

However, the calculations of minor loops should start with a demagnetized state of the film that is not known *a priori*. To find it we apply a "numerical cooling" procedure to our system. It is the solution of Landau-Lifshitz-Gilbert (LLG) equation with additional noise.

$$\frac{d\varphi(\boldsymbol{r})}{dt} = -\gamma\lambda|\widetilde{\boldsymbol{h}}|\sin(\varphi(\boldsymbol{r}) - \varphi_h) + \zeta(\boldsymbol{r}, t) \quad (S3)$$

Here $\varphi(\boldsymbol{r})$ is the polar angle describing the direction of magnetization. $\widetilde{\boldsymbol{h}}$ is the dimensionless effective field $\widetilde{\boldsymbol{h}} = (\partial \mathcal{F}/\partial \boldsymbol{M})/M_0$ and $\varphi_h$ is its polar angle. $\gamma$ is the gyromagnetic ratio and $\lambda$ is damping parameter. $\zeta(\boldsymbol{r}, t)$ is the random force with the correlation function

$$\langle\zeta(\boldsymbol{r}, t)\zeta(\boldsymbol{r}', t')\rangle = \gamma\lambda T_N \delta(t - t')\delta(\boldsymbol{r} - \boldsymbol{r}') \quad (S4)$$

where $T_N = T/M_0\mu_{Co}$ is the dimensionless "numerical temperature", $T$ is temperature and $\mu_{Co}$ is the magnetic moment of a single Co atom. We simulate the stochastic equation (eq4) with the temperature gradually decreasing from the value $T_N = 3K_R$ to zero. The resulting "pseudo-ground" states are shown in **Fig. S7** for different parameters. It shows the crossover from magnetic domains and domain walls to CFG. They are used as the initial states for calculations of minor loops.

Although the general picture of zero-magnetization states (domains or spin glass) is robust against modification of the cooling procedure, the typical sizes of domains or of the spin-glass features depend on the time of the numerical cooling $\tau_{c0}$. In Fig. S8 we show how the calculated state depends on this parameter. The value $\gamma\lambda\tau_{c0} = 10^3$ is used for the calculation of the minor loops.

To calculate the minor loops the adiabatic modification of field $H$ starts from the demagnetized state. The magnetic field then is changed from zero to some value $H_{max}$. Three loops between $H_{max}$ and $-H_{max}$ are made to saturate the magnetization dependence on the magnetic field and then the minor loop is recorded. The result is averaged over 500 disordered numerical samples. Then the minor loop is centered and coercive field $H_C^*$ and remnant magnetization $M_R^*$ are recorded. This procedure is shown in **Fig. S9**. Eqs. (S3) and (S4) allow to calculate the xy-magnetization properties. However, these equations do not allow magnetization to have z-component. Therefore, a vortex in this model are just a singularity in 2D energy density. To study the vortex structure, we apply the vector form of LLG equations.

$$\frac{d\mathbf{M}}{dt} = -\gamma[\mathbf{M}\times\widetilde{\mathbf{H}}] - \frac{\gamma\lambda}{M_0}\Big[\mathbf{M}\times[\mathbf{M}\times\widetilde{\mathbf{H}}]\Big]. \quad (S5)$$

Here the Langevin forces are included in the effective magnetic field $\widetilde{\mathbf{H}} = \partial\mathcal{F}/\partial\mathbf{M} + \mathbf{H}_\xi$, where the cartesian components $H_\xi^{(\alpha)}$ of $\mathbf{H}_\xi$ have the following correlation function

$$\langle H_\xi^{(\alpha)}(\mathbf{r},t)H_\xi^{(\beta)}(\mathbf{r}',t')\rangle = \frac{2T}{\gamma\lambda M_0}\delta_{\alpha\beta}\delta(t-t')\delta(\mathbf{r}-\mathbf{r}'). \quad (S6)$$

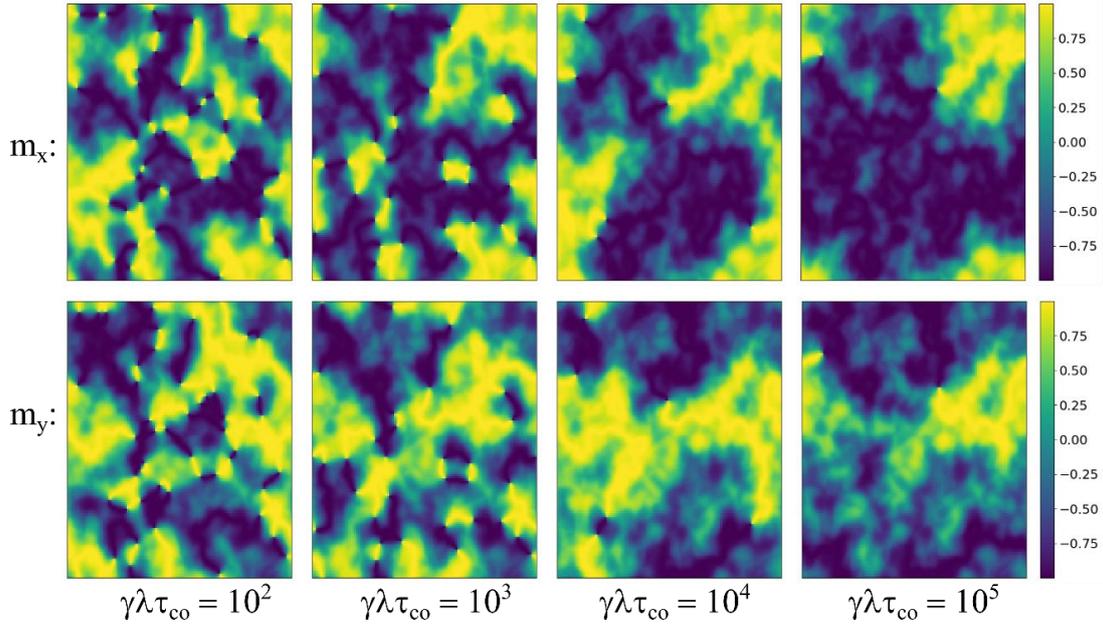

**Fig. S8**: *dependence of the demagnetized state on the time of numerical cooling. All the states are calculated for $r_c = 2\Delta$ and $K_U = 0.02 K_R$.*

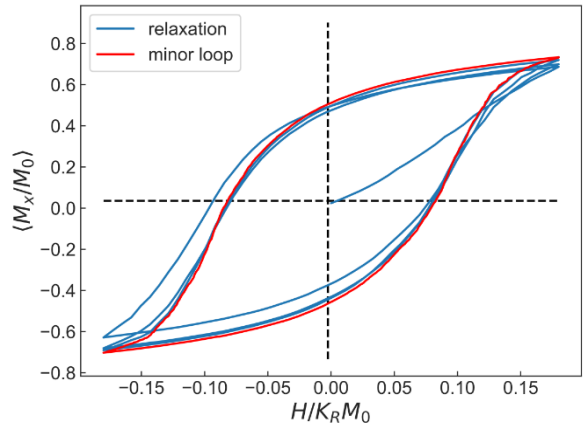

**Fig. S9** *Minor loop calculation.*

To verify the validity of our model we have quantitatively compared the hysteresis loops at 150 K for Co/Al and Co/C$_{60}$ with the micromagnetic model loops. Such comparison is reported in **Fig. S10**. For the experimental values of both samples we have rescaled the applied field by the anisotropy field 75 mT, estimated by looking at the closure field of the Co/C$_{60}$ loop (i.e. the field for which the forward and backward branches are splitting). As can be seen from **Fig. S10**, The Co/C$_{60}$ hysteresis loops is reasonably well reproduced by a K$_R$/K$_U$ ratio of 50 and a correlation radius

$r_c=4\Delta$. Noteworthily, the Co/Al hysteresis loop is reproduced by the theoretical loops considering $r_c=0.2\Delta$, that is a correlation radius that is 5 times less than the re-normalized magnetic length.

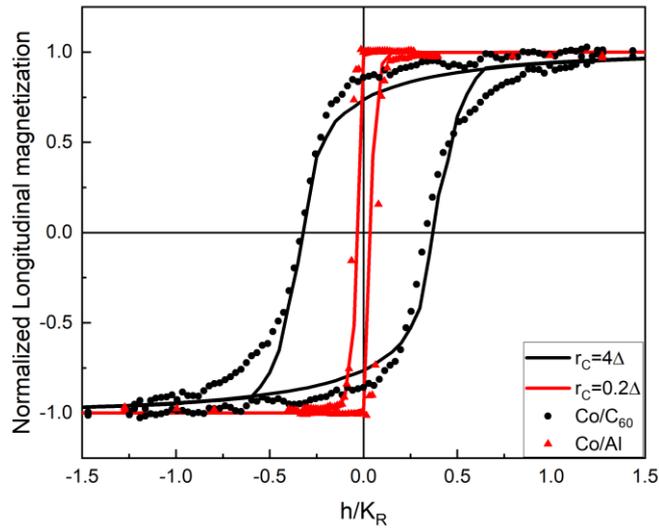

**Fig. S10** *Comparison between experimental hysteresis loops for Co/$C_{60}$ (black points) and Co/Al (red points) and micromagnetic simulations considering an anisotropy ratio $K_U/K_R=0.02$ and magnetic length ratios $r_c/\Delta=4$ (black line) and $r_c/\Delta=0.2$. Experimental data measured at T=150K. Applied field is rescaled for both Co/Al and Co/$C_{60}$ by the closure field of the latter.*

## S. V MAGNETIC FORCE MICROSCOPY CHARACTERIZATION

MFM measurements were taken at RT in an applied field of 2 mT, applied after a full magnetization reversal cycle with an opposite field of -60 mT. **Fig. S11** shows both the topographic and the magnetic contrast map on the Co(5nm)/$C_{60}$(25nm), highlighting how the magnetic signal is not affected by any morphological defect. For completeness we also report the enlarged area from which the data reported in Fig.5 e of the main text are taken. We also add 5 other profile along 5 different vortexes to highlight the magnetic signal depth.

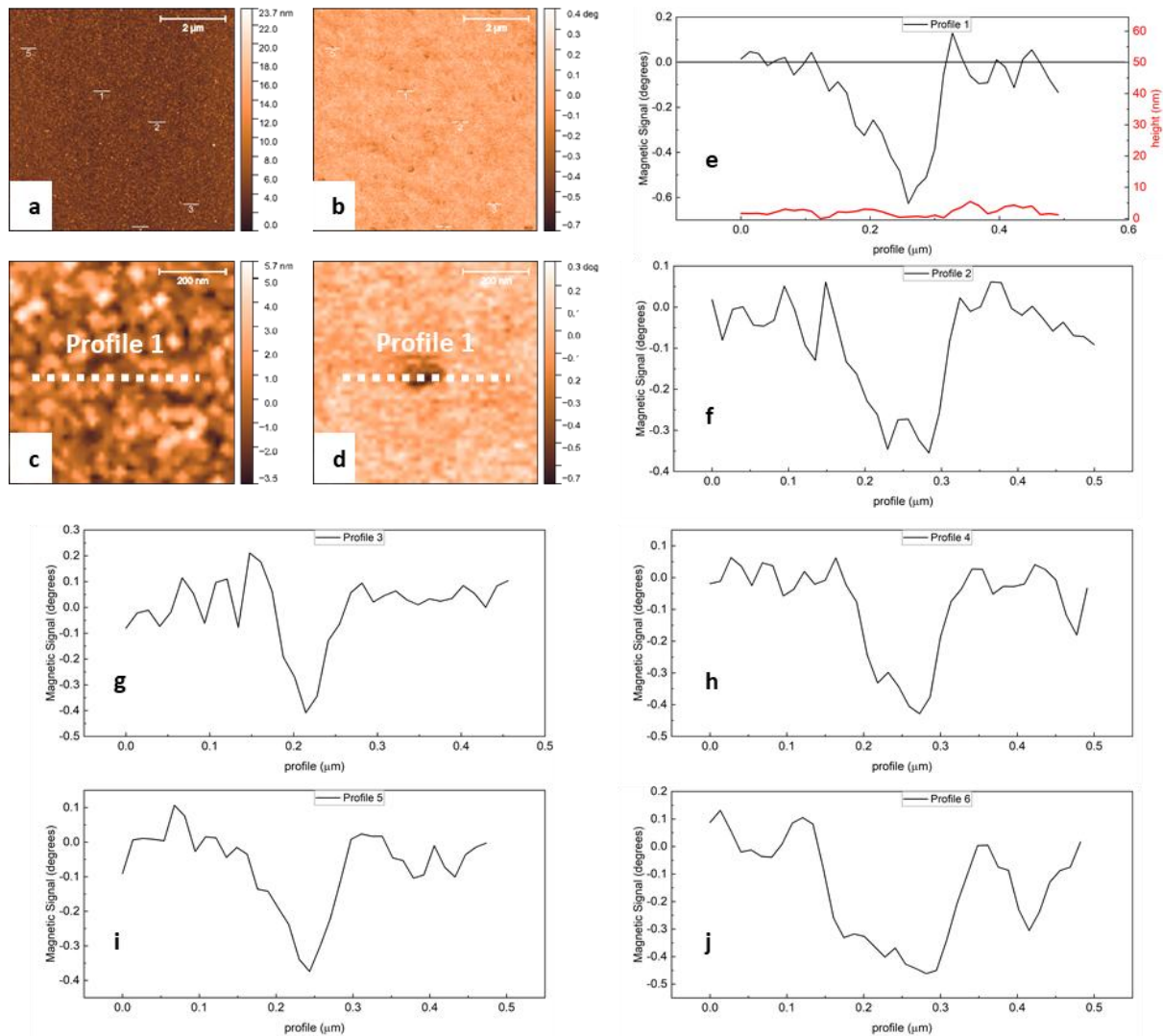

**Fig. S11** *a Topographic and b Magnetic signal maps of the Co/C$_{60}$ sample. The area around Profile 1 (the one reported in **Fig.5 e** of the main text) is reported in c and d respectively. e-j report the 6 profiles of the magnetic signal taken along different magnetic vortexes. e also reports the corresponding topography profile with y-axis adjusted to highlight the 50 nm height (w.r.t. the surface) at which the Magnetic signal was taken.*

## S. VI RT and raw T=150 K hysteresis loops

In **Fig. S12 a** we report the hysteresis loops of Co(5nm) samples at RT, showing an increase in coercivity accompanied by enlarged loop area. For completeness, we report also the non-symmetrized version of the loops shown in **Fig. 1a** of the main article in **Fig. S12 b**, along with the other Co/C$_{60}$ and Co/Gaq$_3$ loops used to extract the coercivities reported in **Fig. 5** of the main text. Of particular interest are the Co/Gaq$_3$ loops, showing (for the loop reported in **Fig. 1a**) no clear double jump but a "peak" at μ$_0$H=0 mT that is attributed to an artifact of the measurement. In fact, by inspecting the raw hysteresis loops used for the averaged Co/Gaq$_3$ loop at 150 reported in **Fig. 1a** of the main article, a spurious oscillating signal is present in the voltage measured by the MOKE photodetector. Its presence affected the average loop resulting in an artificial kink around H=0. Note moreover that the other loop obtained for a different Co/Gaq$_3$ sample does not show any double phase behaviour of the magnetization reversal.

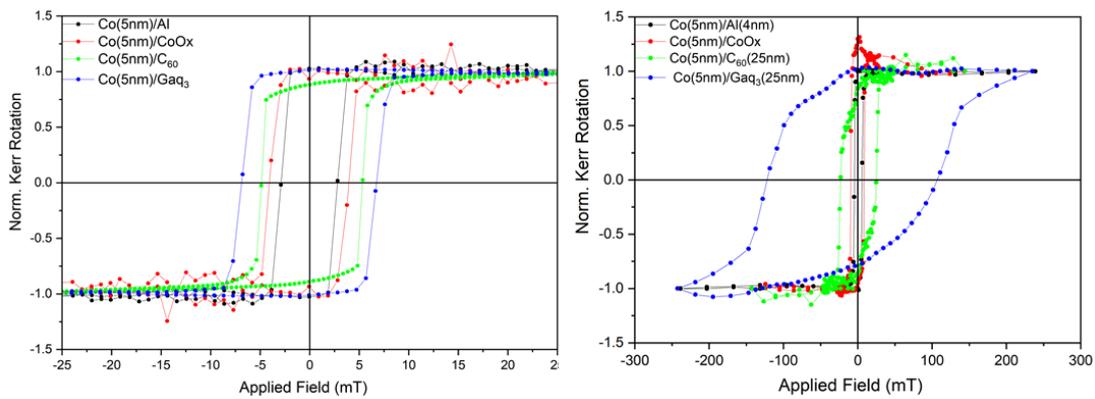

**Fig. S12 a** *RT hysteresis loops of Co(5nm) samples.* **b** *non-symmetrized, normalized hysteresis loops of Co(5nm) samples measured at T=150K (same data shown in Fig. 1)*

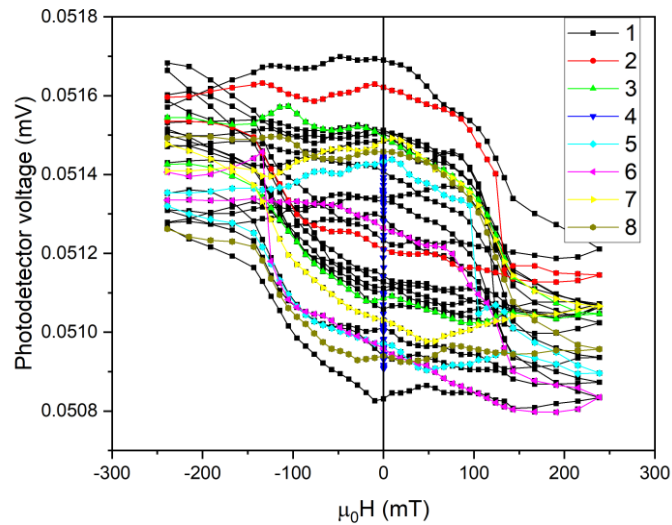

**Fig S13** raw data of the Photodetector output voltage vs applied magnetic field for each averaged loops in the final Co/Gaq$_3$ loop measured at 150 K. A clear oscillating spurious signal is present, affecting the true shape of the hysteresis loop of the sample. Its inclusion in the average process gave rise to the kink appearing in the return branch (H+ to H-)of the hysteresis loop reported in **Fig. S12 b**.

**S. VII GRAIN SIZE and RT, 100 K coercivity for 3,5 and 7 nm thick Co layers**

We report in Table S1 the valiues of the mean grain size for 3, 5 an 7 nm thick Co layers. They were extracted by the procedure reported in Rasigni *et al.*[3], by fitting the 1D fast-axis autocorrelation function of 1 µm Co surfaces, obtained by ex-situ AFM at RT. The estimated grain sizes are compatible, for different thicknesses, within the error bars. We also report in Table S1 the values of the coercivity of Co/$C_{60}$ samples at RT and 80 K.

| Co thickness t (nm) | 3 | 5 | 7 |
|---|---|---|---|
| **Mean grain size (nm)** | 9 ± 1 | 11 ± 1 | 11 ± 1 |
| **RT Co(t)/$C_{60}$ µ₀H$_C$ (mT)** | 6.5 ± 0.4 | 4.7 ± 0.3 | 3.2 ± 0.4 |
| **RT Co(t)/Al µ₀H$_C$ (mT)** | 4.6 ± 0.4 | 3.6 ± 0.3 | 3.1± 0.4 |
| **80K Co(t)/$C_{60}$ µ₀H$_C$ (mT)** | 227 ± 25 | 84 ± 5 | 9.6 ± 0.4 |
| **80K Co(t)/Al µ₀H$_C$ (mT)** | 9.5 ± 0.5 | 4.8 ± 0.4 | 5.0 ± 0.3 |

*Table S1*: *Mean Grain size of Co layer and coercivities of Co/$C_{60}$ samples as a function of the Co thickness.*

**References (supplementary)**